\documentclass[prb,aps,twocolumn,showpacs]{revtex4-1}%
\usepackage{amsmath}
\usepackage{amsfonts}
\usepackage{amssymb}
\usepackage{graphicx}%
\usepackage{bm}
\begin{document}
\title{Is the term ``type-1.5 superconductivity" warranted by Ginzburg-Landau theory? }
\author{V. G. Kogan, J. Schmalian}
\affiliation{Ames Laboratory and Department of Physics \& Astronomy, Iowa State University, Ames, IA 50011}
 
\date{Aug.2, 2010  }
 
\begin{abstract}

 It is shown that within the Ginzburg-Landau (GL) approximation   the order parameters $\Delta_1(\bm r, T)$ and $\Delta_2(\bm r, T)$ in two-band superconductors vary on 
  the same  length scale, the difference in zero-$T$ coherence lengths $\xi_{0\nu}\sim\hbar v_{F\nu}/\Delta_\nu(0)$, $\nu=1,2$   notwithstanding.  
 This amounts to a single physical GL parameter $\kappa$ and the classic GL dichotomy: $\kappa<1/\sqrt{2}$ for type-I and  $\kappa>1/\sqrt{2}$ for type-II. 
\end{abstract}

\maketitle

\section{Introduction}

  The physics of all superconductors near their
critical temperature, $T_{c}$, is based on the  GL  theory.   
\cite{Ginzburg50} This includes multi-band superconductors with distinct
sheets of the Fermi surface. 
A number of recent papers deal with two-band materials with coefficients of the GL free energy (for the field-free state),
  \begin{eqnarray}
  F   =  \sum_{\nu=1,2}\left(a_\nu \Delta_\nu^2  +  
    b_\nu \Delta_\nu^4  /2 \right) -  2 \gamma  \Delta_1\Delta_2 \,,
\label{energy-2b} 
\end{eqnarray}
 introduced phenomenologically, see, e.g., Ref.\,\onlinecite{Zade}.  Choosing these coefficients in various ways, one could arrive to a number of the choice dependent  conclusions. \cite{Babaev,Babaev_last}     However, the coefficients can  be derived from  microscopic theory; they are certain functions  of the microscopic coupling constants responsible for superconductivity and  of temperature $T$. This has been done time ago by Tilley \cite{Tilley} and later by  Zhitomirsky and Dao  \cite{Zhit} who have shown, within a weak-coupling model, that the coefficients $a_\nu$ do not have the familiar GL form $\alpha(T-T_c)$. Instead, they acquire a constant part, $const+\alpha(T-T_c)$, which is intimately related to the constant $\gamma$ of the mixed Josephson-type term to ensure  $\Delta_\nu\propto \sqrt{T_c-T}$  near $T_c$. 

 We argue in this work that the ratio of the order parameters is $T$ independent in the GL domain, 
\begin{equation}
\Delta_1(\bm r, T)/\Delta_2(\bm r, T)=const\,,
\label{main}
\end{equation}
 with the constant depending on   interactions responsible for superconductivity. 
 The one-dimensional version of Eq.\,(\ref{main})   has first been obtained while solving the GL problem of the  interface energy  between superconducting and normal phases relevant for the distinction between type-I and type-II two-band superconductors. \cite{Jani}   For the strong intraband scattering (the dirty limit), the result (\ref{main}) has been obtained by Koshelev and Golubov provided  the interband scattering could be disregarded.\cite{Kosh-Gol} 
 Here, we establish this result for {\it any} problem in the GL domain. 

We show that   the equations for  $\Delta_1(\bm r, T)$ and $ \Delta_2(\bm r, T)$ are reduced to one independent GL equation.    In other words, there is a single complex order parameter describing the two-band superconductor in the GL domain and, as a consequence, a single length scale $\xi$ for   spatial variation of  both $\Delta_1(\bm r, T)$ and $ \Delta_2(\bm r, T)$.  

Our results, along with the earlier critique  \cite{Sasha}  and a comprehensive review by Brandt and Das,  \cite{Brandt-Das}    question  validity of  publications  discussing properties of MgB$_2$  within the GL framework where each band is attributed with its own coherence length 
and sometimes even with its own penetration depth, see, e.g., Ref.\,\onlinecite{Mosch} and references therein. 

We   stress   that our claim that the gap functions $\Delta_\nu(\bm r, T)$     change on the same length scale relates exclusively to the temperature domain, however narrow it could be, where the GL theory is valid. Out of this domain and at low temperatures in particular, different  length scales $\sim\hbar v_{F\nu}/\Delta_\nu(0)$ may enter and result in properties substantially different from those in the GL region. Still, as long as the GL energy functional is used, the assumption of two coherence lengths cannot be justified.


Below, we   discuss   the phenomenologic two-band GL theory and  later confirm our conclusions within a weak-coupling microscopic scheme.\\

  \section{Two-band GL in field}

The   two-band GL functional   reads:  
\begin{eqnarray}
{\cal   F}  &=& \int dV \Big\{ \sum_{\nu=1,2}\left(a_\nu | \Delta_\nu |^2  
 + \frac{b_\nu}{2}| \Delta_\nu|^4 + K_\nu |\bm \Pi\Delta_\nu |^2\right)   \nonumber\\
 & -&   \gamma \left( \Delta_1\Delta_2^*+\Delta_2\Delta_1^*\right)+\frac{ B ^2}{8\pi } \Big \}\,.
\label{Gibbs} 
\end{eqnarray}
 where 
${\bm \Pi}={\bm \nabla}  + 2\pi i {\bm A}/ \phi_0$ and the constant $\gamma$ along with the coefficients $a,b,K$  will be given later.
The GL equations are minimum conditions for the functional (\ref{Gibbs}). One obtains  varying ${\cal F}$ with respect to $\Delta_\nu^*$:
 \begin{eqnarray}
a_1  \Delta_1  +b_1 \Delta_1|\Delta_1|^2 - \gamma \Delta_2 - K_1 \bm\Pi^2\Delta_1 =0\,, \label{GL1d}\\
a_2  \Delta_2 +b_2 \Delta_2|\Delta_2|^2- \gamma \Delta_1 - K_2\bm\Pi^2\Delta_2 =0\,. \label{GL2d} 
\end{eqnarray}

We now recall    that in the one-band  GL equation,
 \begin{eqnarray}
a  \Delta   +b  \Delta |\Delta |^2   - K  \bm\Pi^2\Delta  =0\,, \label{GL-standard} 
\end{eqnarray}
all terms are of the same order $(1-T/T_c)^{3/2} =\tau^{3/2}$  ($\Delta\propto\tau^{1/2}$, $a\propto\tau$, and $\Pi^2\propto\xi^{-2}\propto\tau$). This is not so for Eqs.\,(\ref{GL1d}),    (\ref{GL2d}) because $\gamma$ is a constant and $a_\nu$ may contain  constant parts. 

Having this in mind, we express $\Delta_2$ in terms of $\Delta_1$ from Eq.\,(\ref{GL1d}) and substitute the result in   Eq.\,(\ref{GL2d}) keeping only terms up to the order $\tau^{3/2}$:
 \begin{eqnarray}
(a_1a_2-\gamma^2)  \Delta_1  &+&(b_1a_2+b_2a_1^3/\gamma^2) \Delta_1|\Delta_1|^2  \nonumber\\
& -& (a_1K_2+a_2K_1)  \bm\Pi^2\Delta_1 =0\,. \label{GL1f}
\end{eqnarray}
Similarly, one obtains an equation for $\Delta_2$:
 \begin{eqnarray}
(a_1a_2-\gamma^2)  \Delta_2  &+&(b_2a_1+b_1a_2^3/\gamma^2) \Delta_2|\Delta_2|^2  \nonumber\\
& -& (a_1K_2+a_2K_1)  \bm\Pi^2\Delta_2=0\,.\label{GL2f}
\end{eqnarray} 
In zero field, one has $\Delta_\nu^2 \propto (a_1a_2-\gamma^2)$, so that at $T_c$,
$ a_1a_2-\gamma^2 =0$,
and therefore $a_\nu$ must contain constant parts,
 \begin{eqnarray}
 a_\nu= a_{ \nu\,c}- \alpha_\nu\tau\,,\label{a's}
\end{eqnarray} 
such that $ a_{1c}a_{2c}=\gamma^2$. 

Eqs.\,(\ref{GL1f}) and (\ref{GL2f}) for $\Delta_\nu$ can now be written as:
 \begin{eqnarray}
-\alpha\tau  \Delta_1 +\beta_1 \Delta_1|\Delta_1|^2 
-K  \bm\Pi^2\Delta_1 =0\,, \label{GL1ff}\\  
 -\alpha\tau  \Delta_2 +\beta_2 \Delta_2|\Delta_2|^2 
 -K  \bm\Pi^2\Delta_2 =0\,, \label{GL2ff}
\end{eqnarray}
with 
 \begin{eqnarray}
\alpha&=&\alpha_1+\alpha_2\,, \qquad\qquad K = a_{1c}K_2+a_{2c}K_1\,,\nonumber\\
 \beta_1&=&b_1a_{2c}+b_2a_{1c}^3/\gamma^2\,,\quad  \beta_2=b_2a_{1c}+b_1a_{2c}^3/\gamma^2\,.\quad
\end{eqnarray}
We note that within the accuracy of the GL theory, up to ${\cal O}(\tau^{3/2})$, the equations for $\Delta_1$ and $\Delta_2$ are coupled only via the vector potential.

 In particular, in zero field we have
\begin{eqnarray}
  \Delta_{\nu\,0}^2 =  \alpha \tau /\beta_\nu  \,, 
\label{Delta's} 
\end{eqnarray}
so that the ratio
\begin{eqnarray}
 \frac{ \Delta_{10}^2(T)}{\Delta^2_{20}(T)} = \frac{\beta_2}{\beta_1}  \,, 
\label{Delta's} 
\end{eqnarray}
comes out to be $T$ independent in the GL domain.
 
Furthermore, one easily checks that for any solution $\Delta_1(\bm r,T)$ of Eq.\,(\ref{GL1ff}), Eq.\,(\ref{GL2ff}) is satisfied by 
  \begin{eqnarray}
  \Delta_2(\bm r,T) = \Delta_1(\bm r,T) \sqrt{\beta_1/\beta_2 } \,. \label{D2} 
\end{eqnarray}
In particular, this  implies  that in equilibrium $\Delta_1(\bm r,T)$ and $\Delta_2(\bm r,T)$ must have either the same phases or the phases differing by $\pi$. \cite{Gurevich}    
It is found in Ref.\,\onlinecite{Zhit}  that for small $\gamma$ the ratio $\Delta_2/\Delta_1$ changes away of $T_c$; we note, however, that this deviation is beyond the GL accuracy. Reliable results beyond GL can be obtained only within microscopic approaches like Gor'kov or Bogolyubov - de Gennes theories.
 
Moreover, introducing  the order parameters 
normalized on their zero-field values, 
\begin{eqnarray}
 \frac{ \Delta_1}{ \Delta_{10}(T) }= \frac{ \Delta_2}{ \Delta_{20}(T) }=\Psi  \,, 
\label{dimensionless} 
\end{eqnarray}
both Eq.\,(\ref{GL1ff}) and (\ref{GL2ff}) are reduced to one:
 \begin{eqnarray}
  \Psi (1-  |\Psi|^2)=-\frac{K}{\alpha\tau} \bm\Pi^2\Psi  \,. \label{GL00}
\end{eqnarray}
Thus,  the length scale of the space variation of both $\Delta_1$ and $\Delta_2$,  the coherence length, is given by
 \begin{eqnarray}
\xi^2 =  K/\alpha\tau   \,. \label{xi}
\end{eqnarray}


\section {Microscopic weak-coupling two-band model near $T_c$.} 

To establish connection of GL equations with the two-band microscopic theory we turn to a weak-coupling model for clean and isotropic materials (not because these restrictions  are unavoidable, but rather due to the model simplicity).   

Perhaps, the simplest formally weak-coupling approach is based on
the Eilenberger quasiclassical formulation of the Gor'kov equations 
 valid for general anisotropic order parameters and  Fermi 
surfaces. \cite{E}   Eilenberger functions $f,g$ for clean materials 
in zero-field  obey the system:
\begin{eqnarray}
0&=& \Delta g  - \hbar\omega f \,,\label{eil1}\\
g^2&=&1-f^2\,, \label{eil3}\\
\Delta( {\bm k})&=&2\pi TN(0) \sum_{\omega >0}^{\omega_D} \Big\langle
V({\bm k},{\bm k}^{\prime\,}) f({\bm k}^{\prime},\omega)\Big\rangle_{{\bm k}^{\prime\,}}.
   \label{self-cons}
\end{eqnarray}
Here, ${\bm k}$ is the Fermi momentum; $\Delta$ is the order parameter that may depend on the position ${\bm  k}$ at the Fermi surface.    Further,
$N(0)$ is the total density of states (DoS) at the Fermi level per spin;
the   Matsubara frequencies  are given by $\hbar\omega=\pi
T(2n+1)$ with an integer $n$, and $\omega_D$ is the Debye
frequency; $\left\langle...\right\rangle$ stands for averages over
the Fermi surface.

   Consider a model material with the gap given by
\begin{equation}
\Delta ({\bf k})= \Delta_{1,2}\,,\quad {\bf k}\in   F_{1,2} \,, 
 \label{e50} 
\end{equation}
where $F_1,F_2$ are two sheets of the Fermi surface. 
 The gaps are assumed constant at each band. Denoting
 DoS  on the two parts as $N_{1,2}$, we have 
 for a quantity $X$ constant at each Fermi sheet:
 \begin{equation}
\langle X \rangle = (X_1 N_1+X_2 N_2)/N(0) = n_1X_1+
n_2X_2\,,
\label{norm2}
\end{equation}
where $n_{1,2}= N_{1,2}/N(0)$; clearly, $n_1+n_2=1$. 

  Equations (\ref{eil1}) and (\ref{eil3})
 are easily solved:
\begin{equation}
f_\nu  = \Delta_\nu /
\beta_\nu  ,\quad g_\nu   =   \hbar\omega/
\beta_\nu  ,\quad
\beta_\nu ^2=\Delta_\nu ^2+ \hbar^2\omega^2\,, \label{f_n}
\end{equation}
where  $\nu=1,2$ is the band index. 
The  self-consistency equation (\ref{self-cons}) takes the form:
\begin{eqnarray}
\Delta_\nu=  \sum_{\mu=1,2} n_\mu \lambda_{\nu\mu} 
\Delta_\mu \sum_{\omega }^{\omega_D}\frac{2\pi T}{\beta_\mu} ,\quad
   \label{self-cons1}
\end{eqnarray}
where $\lambda_{\nu\mu} = N(0)V_{\nu\mu}$ are dimensionless 
effective interaction constants.  The notation commonly
used   in literature,   $\lambda^{(lit)}_{\nu\mu}=n_\mu \lambda_{\nu\mu}$, includes DoS'. We find our notation convenient since, being related to the coupling potential, our coupling matrix is symmetric: $\lambda_{\nu\mu}=\lambda_{\mu\nu}$.  
 

It is seen from the system (\ref{self-cons1})  that   $\Delta_{1,2} $   turn zero at the same temperature $T_c$ unless $\lambda_{12}=0$ and equations (\ref{self-cons1})  decouple, the property that has   been noted in earlier work. \cite{M59,SMW,BTG}  
As $T\to T_c$,   $\Delta_{1,2}\to 0$, and $\beta\to\hbar\omega$. 
The sum over $\omega$ in Eq.\,(\ref{self-cons1}) is readily evaluated:
\begin{eqnarray}
S=  \sum_{\omega }^{\omega_D}\frac{2\pi T}{\hbar\omega}\Big|_{T_c} = \ln\frac{2\hbar\omega_D}{T_c\pi e^{-\gamma}}=  \ln\frac{2\hbar\omega_D}{1.76\,T_c }, 
   \label{S}
\end{eqnarray}
 $\gamma=0.577$ is the Euler constant.  This relation can also be written as
 \begin{eqnarray}
1.76\,T_c=  2\hbar\omega_D e^{-S}\,. 
   \label{S-Tc}
\end{eqnarray}
 The system (\ref{self-cons1}) at $T_c$ is linear and homogeneous:
\begin{eqnarray}
\Delta_1&=&  S (n_1 \lambda_{11} \Delta_1+ n_2 \lambda_{12} \Delta_2)\,,\nonumber\\
\Delta_2&=&  S (n_1 \lambda_{12} \Delta_1+ n_2 \lambda_{22} \Delta_2) .
   \label{systemTc}
\end{eqnarray}
The zero   determinant   gives $S$ and, therefore, $T_c$:
\begin{eqnarray}
  S^2 n_1n_2\eta &-& S (n_1 \lambda_{11}   + n_2 \lambda_{22})+1=0 , \label{det=0}\\
  \eta& =&  \lambda_{11}  \lambda_{22}-\lambda_{12}^2\,.
 \label{eta}
 \end{eqnarray}
The roots of this equation  are:
\begin{eqnarray}
  S=  \frac{ n_1 \lambda_{11}   + n_2 \lambda_{22} \pm\sqrt{ (n_1 \lambda_{11} - n_2 \lambda_{22})^2 +4n_1n_2\lambda^2_{12}}}{2n_1n_2\eta}.\nonumber \\
 \label{S1} 
\end{eqnarray}
Various possibilities that arise depending on  values of $\lambda_{\mu\nu}$ are discussed, e.g., in Refs.\,\onlinecite{M59}-\onlinecite{gamma}. 
Introducing  $T$-independent quantities,
\begin{eqnarray}
 S_1 =\lambda_{22}-n_1\eta S\,,\quad   S_2 =\lambda_{11}-n_2\eta S\,,
\label{S12} 
\end{eqnarray}
we write Eq.\,(\ref{det=0}) as
 \begin{eqnarray}
 S_1  S_2 =\lambda_{12}^2\,,
\label{S1S2} 
\end{eqnarray}
 the form useful  for manipulations below.

 If   $ \lambda_{12}=0$,   Eq.\,(\ref{S1})
provides two roots $ 1/n_1\lambda_{11}$ and  $ 1/n_2\lambda_{22}$.
The smallest one gives $T_c$, whereas the other corresponds to
the temperature at which the second gap turns zero. We note that this situation is unlikely; it implies that the ever present Coulomb repulsion is exactly compensated by the effective interband  attraction.

Since the determinant of the 
system (\ref{systemTc}) is zero, the two equations are 
equivalent and  give at $T_c$: 
\begin{eqnarray}
\left(\frac{\Delta_2}{\Delta_1}\right)_{T_c}= \frac{ 1-n_1\lambda_{11}S}{n_2\lambda_{12}S} 
  \,.   \label{sign}
\end{eqnarray}
When the right-hand side is negative, $\Delta$'s are of opposite 
signs.  
 Within the  one-band BCS, the sign of $\Delta $ is a matter of convenience; for 
 two bands, $\Delta_1$ and $\Delta_2$ may have equal or opposite signs. \cite{mazin-golubov}  
 
After simple algebra, Eq.\,(\ref{sign}) can be manipulated to 
\begin{eqnarray}
\left(\frac{\Delta_2 }{\Delta_1}\right)^2 _{T_c}= \frac{S_{1}}{S_2 } 
  \,.   \label{ratio2}
\end{eqnarray}
We thus obtain by comparing with Eq.\,(\ref{Delta's}) or (\ref{D2}) the ratio of phenomenological coefficients in terms of microscopic couplings: $\beta_1/\beta_2=S_1/S_2$. 
 We have seen above that within the GL approximation this ratio remains the same at any $T$ in the  GL domain     not only for a uniform field-free state (or for $\gamma\to\infty$ as in Ref.\,\onlinecite{Peeters}) but for any situation with $\Delta$'s depending on coordinates in the presence of magnetic fields.  
 
 We note that the proportionality of 
$\Delta_1$ and $\Delta_2$ has also been shown to hold within microscopic weak-coupling theory in the dirty limit by Koshelev and Golubov.\cite{Kosh-Gol} It is also worth mentioning here that the above proof of this  proportionality based on the GL approach is quite general and holds for any scattering, gap anisotropies etc.

 In the following we use the GL coefficients obtained in Refs.\,\onlinecite{Tilley} and \onlinecite{Zhit}. In our notation they read:
\begin{eqnarray}
a_\nu&=&\frac{N(0)}{\eta}\,(S_\nu-n_\nu\eta\tau), \,\,\,\, 
  b_\nu = \frac{N(0)}{W^2}\,n_\nu,\,\,\,\, W^2=\frac{8\pi^2T_c^2}{7\zeta(3)}, \nonumber\\
   \gamma &=&\frac{N(0)}{\eta}\,\lambda_{12}\,,\qquad   K_\nu =\frac{N(0) \hbar^2v_\nu^2}{6W^2} \,n_\nu\,,
\label{b,gam}
\end{eqnarray}
where the energy scale $W\sim\pi T_c$ is introduced for brevity and  $v_\nu $ are the Fermi velocities in two bands which  for simplicity are assumed isotropic. 
We, in fact,   confirmed Eqs.\,(\ref{b,gam}) of Zhitomirsky and Dao   employing different methods  (except our 
$b_\nu $   is by a factor of 2 larger than that of  Ref.\,\onlinecite{Zhit}).
It is worth noting  that the microscopically derived $a_\nu$ are not proportional to $\tau$ as in the standard one-band GL  unless one of the parameters $S_\nu$ is zero; given the condition (\ref{S1S2}) this may happen only if $\lambda_{12}=0$.   
This feature of the two-band GL is sometimes overlooked. \cite{India,Lin}    

As     stressed in Ref.\,\onlinecite{Zhit}, the term   $K_\nu |\bm \Pi\Delta_\nu |^2$ with  order parameters gradients is the only possible in the GL energy, although the symmetry may allow for other combinations of   gradients.

The coefficients entering the GL Eqs.\,(\ref{GL1ff}) and  (\ref{GL2ff}) are:
\begin{eqnarray}
\alpha= \frac{N(0)^2C}{\eta}  ,\,\,\,  K= \frac{\hbar^2\tilde{v}^2 N(0)^2}{6W^2\eta},\,\,\, 
\beta_\nu = \frac{N(0)^2D\,S_\nu}{\eta W^2\lambda_{12}^2} ,\qquad
\label{coeff}
\end{eqnarray}
where 
   \begin{eqnarray}
\tilde{v}^2=n_1S_2  v^2_1 +n_2S_1  v^2_2 \, \label{v-bar1}
\end{eqnarray}
has the dimension of a squared velocity and 
 \begin{eqnarray}
C= n_1S_2 +n_2S_1\,,\quad D=n_1S_2^2 +n_2S_1^2\,\label{C,D}
\end{eqnarray}
are constants. 

Hence, we can express the length scale (\ref{xi}) of the space variation of both $\Delta_1$ and $\Delta_2$ in the GL domain    in terms of microscopic parameters:
 \begin{eqnarray}
\xi^2 =  \frac{\hbar^2 \tilde{v}^2}{2W^2C\tau}  \,. \label{xi-micro}
\end{eqnarray}
The upper critical field follows: $H_{c2}=\phi_0/2\pi\xi^2$. 
 The one-band limit is obtained by setting $n_1=1$, $n_2=0$ so that $C=S_2$    and $\tilde{v}^2=  S_2v^2/3$ that yields   $\xi^{2} = 7\zeta(3)\hbar^2v^2/48\pi^2T_c^2\tau$ as it should.

 
  Variation of the free energy ${\cal   F}$ with respect to the vector potential $\bm A$ gives the current density. 
Following the standard procedure we obtain for the {\it penetration depth} of a weak magnetic field: 
\begin{eqnarray}
\frac{1}{ \lambda^2}  = \frac{32\pi^3  }{\phi_0^2}   \sum_{\nu=1,2}    \Delta_{\nu 0}^2      K_\nu    =\frac{16\pi CN(0)e^2\tilde{v}^2}{c^2D }\,\tau.\quad
\label{lam} 
\end{eqnarray}
 In the one-band limit  this yields the correct result:  $\lambda^{-2} = (16\pi e^2N(0)v^2/3c^2)\tau$. 
 
  A straightforward calculation yields the equilibrium zero-field free energy:
\begin{eqnarray}
F_0= -  N(0)W^2\frac{C ^2}{2D}  \,\tau^2\,.
\label{Fo} 
\end{eqnarray}
 The thermodynamic field $ H_c$ follows: $H_c^2/8\pi = -F_0$. One can show  that the relative specific heat jump at $T_c$ differs from the  one-band value $12/7\zeta(3)=1.43$ by a factor $C^2/D<1$.  \cite{Mosk2}

One can now form the dimensionless GL parameter,
\begin{eqnarray}
\kappa^2 =\frac{\lambda^2}{\xi^2}   =  \frac{c^2W^2D }{8\pi  N(0)e^2\hbar^2 \tilde{v}^4}\,,\quad
\label{lam} 
\end{eqnarray}
and verify the standard relation $H_{c2}/H_c\sqrt{2}=\kappa$.  

 
Finally,  the {\it equilibrium energy}  is  evaluated by 
  substituting  the solutions of the GL equations to the functional (\ref{Gibbs}): 
 \begin{eqnarray}
{\cal   F}=\frac{H_c^2}{4\pi} \int dV \Big\{b^2-\frac{1}{2}|\Psi |^4  \Big \}\,.
\label{F01 } 
\end{eqnarray}
where   $b =  B/H_c\sqrt{2} $ is the dimensionless field. Thus, the   theory of a two-band superconductor near $T_c$ is mapped onto the standard one-order parameter GL scheme.  

In particular, this mapping means that the GL problem of the interface energy   between normal and superconducting phases has the same solution, i.e., $\kappa = 1/\sqrt{2}$ separates type-I and type-II superconductors. This has been demonstrated in Ref.\,\onlinecite{Jani} by solving  numerically the nonlinear system of GL equations (\ref{GL1d}),(\ref{GL2d})  without discarding terms ${\cal O}(\tau^2)$ employed here. 



 \subsection{ Remark on boundary conditions} 

The solution (\ref{D2}) for the two gap functions of the GL Eqs.\,(\ref{GL1ff}) and (\ref{GL2ff}) holds indeed provided the boundary conditions for $\Delta_2$ are the same as for $\Delta_1$ multiplied by the factor $ \sqrt{\beta_1/\beta_2 }$. This is clearly the case for the 1D problem of the S-N interface energy. The same is true for the problem of the single vortex structure: both $\Delta$'s are zero at the vortex center and approach $\Delta_{\nu,0}$ with the correct ratio at infinity. 

However, for, e.g., proximity situations with a two-band superconductor on one side of the contact with a  normal metal, the condition on the superconducting side far from the boundary is satisfied, whereas the question of boundary conditions at the boundary remains open. In this case, one cannot claim that $\Delta(x)$'s are proportional. 
Nevertheless, as is seen from Eqs.\,(\ref{GL1ff}) and (\ref{GL2ff}), the length scale $\xi=\sqrt{K/\alpha\tau}$ is still the same for both order parameters.  



 \section{Discussion}

   Two-band GL equation have been used in a number of  publications where the coefficients in the GL energy functional  $a_\nu, b_\nu, K_\nu$ and $\gamma$ were varied and possible consequences were discussed. Moreover, different $\xi$'s and even $\lambda$'s were assigned to the two bands along with two different $\kappa$'s. This led to speculations that situations may exist where one of the bands  behaves as a type-II superconductor with $\kappa_1>1/\sqrt{2}$, while the other  may   have $\kappa_2<1/\sqrt{2}$ and behave near $T_c$ as the type-I; the superconductivity in such situations was called ``type-1.5".   MgB$_2$ has been suggested as such an example, see, e.g., Ref.\,\onlinecite{Mosch} and references therein.     
  
 The present work  argues that such situations do not exist. The point is that the GL equations are derived from the microscopic theory within certain approximations that lead to the free energy near $T_c$ being  proportional to $(1-T/T_c)^2$ and the 
order parameter (or parameters) varying as $(1-T/T_c)^{1/2}$. Formally, the nonlinear system of GL equations (\ref{GL1d}),(\ref{GL2d}) for  two-band materials can be solved with any accuracy. However, physically there is no point in going to accuracy higher than that of equations themselves; whatever results obtained along these lines will be absolutely unreliable. To get a near-$T_c$ description more accurate than GL, one should go back to microscopic theory that generates many extra terms in the free energy expansion even  for the one-band situation, see, e.g., Ref.\,\onlinecite{Tewordt}, so that the multi-band generalization of such an approach is unlikely to produce a useful theory. It is demonstrated on a one-dimensional problem of Ref.\,\onlinecite{Jani} and is shown for a general case in this paper that {\it within the GL accuracy} both order parameters of a two-band superconductor vary on the same length scale $\xi$ of Eq.\,(\ref{xi}) contrary to requirements of ``1.5 type superconductivity".  

We note that this conclusion holds for the "GL domain" defined as the temperature interval near $T_c$ where the GL expansion can be justified. We do not specify this domain explicitly because its size may vary from one case to another. E.g., it is argued in Ref.\,\onlinecite{Kosh-Gol} that for two dirty bands (with no inter-band scattering) of MgB$_2$, the domain of GL applicability shrinks practically to zero.  However, whatever this size is, within this domain the two order parameters vary on the same length scale. Therefore, attempts to employ the GL functionals  -  on hand -  and  to assume different length scales - on the other - cannot be justified. 
 
Moreover, we show that - within the GL accuracy - the two GL equations for the two-band case are reduced to a single equation for the normalized order parameter; in other words, the two-band superconductor is described by a single complex order parameter. This excludes possibilities of having ``fractional vortices" with exotic properties such as those discussed in Refs.\,\onlinecite{Egor}, \onlinecite{Silaev}.


 Microscopically, our results were derived within a weak coupling theory of clean
superconductors. We believe, however, our conclusions go beyond that. For our results to hold it
is crucial that due to the finite interband Josephson coupling $\gamma $,
the coefficients $a_{\nu}$ in the GL energy remain finite at $T_{c}$. Once
this is guaranteed our qualitative conclusions remain unchanged, even if 
 assumptions of the weak coupling, no-scattering, and isotropy do not apply.\\


 Discussions with R.~Fernandes, J.~Geyer, J.~Clem,  R.~Prozorov, M.~Das, M.~Milosevic,   M.~Zhitomirsky,  A.~Gurevich, R. Mints, and L. Bulaevskii are gratefully appreciated. The work  was
supported by the DOE-Office of Basic Energy Sciences, Division of Materials Sciences and Engineering under Contract No. DE- AC02-07CH11358.


\begin{thebibliography}{99}

\bibitem{Ginzburg50} V. L. Ginzburg and L. D. Landau, Soviet Physics JETP 
\textbf{20}, 1064 (1950).

 \bibitem{Zade}I. N. Askerzade, A. Gencer, and N. G{\"u}cl{\"u}, Supercond. Sci.
Technol. {\bf 15}, L17, (2002).

  
  \bibitem{Babaev}E. Babaev and M. Speight, \prb {\bf 72}, 180502 (2005). 
  
    \bibitem{Babaev_last}E. Babaev and J. Carlstr\"{o}m, arXiv:1007.1965 (2010). 

\bibitem{Tilley}D. R. Tilley, Proc. Phys. Soc. London, {\bf 84}, 573 (1964).
  
 \bibitem{Zhit}M. E. Zhitomirsky and V.-H. Dao, \prb  {\bf 69}, 054508 (2004).
    
\bibitem{Jani} Jani Geyer, Rafael M. Fernandes, V. G. Kogan, and J\"{o}rg Schmalian, \prb {\bf 82}, 104521  (2010).  

\bibitem{Kosh-Gol} A. E. Koshelev and A. A. Golubov, \prl {\bf 92}, 107008  (2004).  


\bibitem{Sasha}A. Gurevich, privat communication.

\bibitem{Brandt-Das} E. H. Brandt and M. P. Das, arXiv:1007.1107 (2010).
 
\bibitem{Mosch}V. Moschalkov {\it et al},
\prl {\bf 102}, 117001 (2009).

 
  \bibitem{Gurevich} This is not the case in the resistive state:  A.~Gurevich and V.~M.~Vinokur,  \prl {\bf 90}, 047004 (2003).
   
\bibitem{E}G.~Eilenberger, Z. Phys. {\bf  214}, 195 (1968).

\bibitem{M59} V.~A.~Moskalenko, Phys. Met. Metallogr. {\bf 8}, 503 (1959).

\bibitem{SMW}H.~Suhl, B.~T.~Matthias and L.~R.~Walker, \prl {\bf 3}, 552 (1959).

\bibitem{BTG}B.~T.~Geilikman, R.~O.~Zaitsev, and V.~Z.~Kresin, Sov. Phys. Solid State, {\bf 9}, 642 (1967).


\bibitem{Maz-Schmal}I.~I.~Mazin and J.~Schmalian,  Phys. C: Supercond. {\bf 469}, 614 (1995).

\bibitem{gamma}V.~G.~Kogan, C. Martin, and R. Prozorov, \prb  {\bf 80},  014507 (2009).

\bibitem{mazin-golubov} A. A. Golubov and I.~I.~Mazin,  Phys. C: Supercond. {\bf 243}, 153 (2009).
  
\bibitem{Peeters} R. Geurts, M. V. Milosevic, and F. M. Peeters, \prb {\bf 81}, 214514 (2010).

 \bibitem{India}M. Karmakar and B. Dey, J. Phys.: Condens. Matter,  {\bf 22},  205701 (2010).

\bibitem{Lin} Shi-Zeng Lin and Xiao Hu, arXive:1007.1940, 2010.  
 
 
 
  \bibitem{Mosk2} V. A. Moskalenko, M. E. Palistrant, V.~M.~Vakalyuk,   Usp. Fiz. Nauk, {\bf 161}(8), 155  (1991).

  \bibitem{Tewordt}L. Neumann and L. Tewordt, Z. Physik {\bf 191}, 73 (1966).
  
   \bibitem{Egor}E. Babaev, \prl {\bf 89}, 067001 (2002). 
   
    \bibitem{Silaev}M. A. Silaev, arXiv:1008.1194 (2010). 
  

\end{thebibliography}
\end{document}